\documentclass[12pt]{article}
\usepackage{cite}
\usepackage{graphicx}
\usepackage{caption}
\usepackage{refstyle}
\usepackage{amsmath}
\usepackage{amsthm}
\usepackage{amsmath}
\usepackage{amsfonts}
\usepackage{amssymb}
\usepackage{bbm}
\usepackage{amsmath,array}
\usepackage{mathtools}
\usepackage{slashbox}
\usepackage[all]{xy}
\usepackage{mathtools}
\usepackage[utf8]{inputenc}
\usepackage[english]{babel}
\usepackage[T3,T1]{fontenc}
\usepackage{fancyhdr}
\DeclareSymbolFont{tipa}{T3}{cmr}{m}{n}
\DeclareMathAccent{\invbreve}{\mathalpha}{tipa}{16}

\usepackage[left=1.3cm,right=1.3cm,top=2.7cm,bottom=2.9cm]{geometry}
\usepackage{setspace}

\title{Optimality of Gaussian in Enlarging HK  Rate Region, and its Overlap with the  Capacity Region of 2-users GIC}

\usepackage{authblk}
\author[1]{Amir K. Khandani\thanks{E\&CE Department, University of Waterloo, Waterloo, Ontario, Canada, khandani@uwaterloo.ca.}}

\begin{document}

\vspace{-5cm}
\maketitle

\vspace{-1cm}
\begin{abstract}
This article\footnote{This article, in some places, refers to weak GIC, but the results are universally applicable to GIC in regimes other than weak. The reason is that the focus of \cite{HK1}, which is a preamble to this work, has been on 2-users weak GIC.} shows that the set of HK constraints correspond to projecting the intersection of two multiple access channels on its sup-spaces. A {\em key property} of HK constraints is that the private message of user 1 (or of user 2) is the last layer in superposition coding for the MAC formed at receiver 2 (or at receiver 1) and will be treated as noise in the decoding operations at receiver 2 (or at receiver 1). This property is used in this article to show that, in a HK rate region based on an additive Gaussian noise model, Gaussian distribution is optimum for enlarging the region.  It is known that the HK rate region is achievable in an interference channel.  On the other hand, reference~\cite{HK1} presents a method for code-book construction in a 2-users weak GIC, using Gaussian inputs, based on covering the boundary of the capacity region in infinitesimal steps. The region constructed in \cite{HK1} coincides with the intersection of the same two multiple access channels that surface in the HK rate region. Reference \cite{HK1} also shows that, due to the aforementioned key property, 2-users GIC capacity region cannot extend beyond the HK rate region.  This means, using Gaussian code-books, the HK rate region is optimally enlarged and the same Gaussian code-books achieve the capacity region of the 2-users GIC.      

\end{abstract}

\begin{figure}[h]
   \centering
   \includegraphics[width=0.6\textwidth]{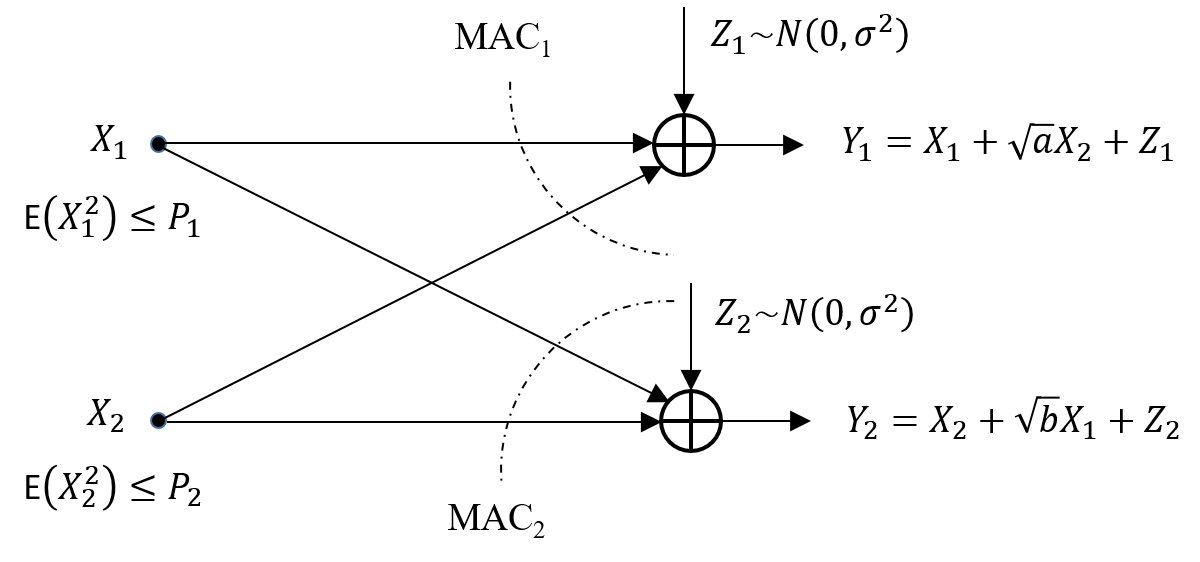}
   \caption{HK constraints \ref{HK1} to  \ref{HK14} in conjunction with additive Gaussian noise model.}
   \label{fig1}
 \end{figure}

This work studies HK achievable rate region  in conjunction with an additive Gaussian noise model shown in Figure~\ref{fig1}.
To derive the HK constraints using notations widely used in the literature,  one should refer to expressions 3.2 to 3.15 on page 51 of~\cite{HK2}, and apply the following changes,
($\mbox{current~article} \leftrightarrow$~\cite{HK2}):
$U_1 \leftrightarrow W_1$, 
$U_2 \leftrightarrow W_2$,
$V_1 \leftrightarrow U_1$, 
$V_2 \leftrightarrow U_2$,
$R_{U_1}\leftrightarrow T_1$, $R_{U_2}\leftrightarrow T_2$, 
$R_{V_1}\leftrightarrow S_1$, $R_{V_2}\leftrightarrow S_2$. Applying these changes, the expanded Han-Kobayashi constraints  and the associated optimization problem  are expressed as follows

\begin{align} 
\label{HK1p}
\mbox{Maximize:}~~~~~ & R_1+\mu R_2  \\
\mbox{Subject to:}~~~ &   \nonumber \\ \label{HK1}
 R_{U_1}    & ~~{\le}~~   I(U_1;Y_1|U_2,V_1)     \\ \label{HK2}
 R_{U_1}  & ~~{\le}~~   I(U_1;Y_2|U_2,V_2)    \\ \label{HK3}
 R_{U_2}   & ~~{\le}~~    I(U_2;Y_1|U_1,V_1)    \\ \label{HK4}
 R_{U_2}   & ~~{\le}~~   I(U_2;Y_2|U_1,V_2)    \\  \label{HK5}
 R_{V_1}  & ~~{\le}~~  I(V_1;Y_1|U_1,U_2)   \\ \label{HK6}
 R_{V_2}  & ~~{\le}~~   I(V_2;Y_2|U_1,U_2)    \\ \label{HK7}  
 R_{U_1}+R_{U_2}   & ~~{\le}~~ I(U_1,U_2;Y_1|V_1)    \\  \label{HK8}
 R_{U_1}+R_{U_2}   & ~~{\le}~~   I(U_1,U_2;Y_2|V_2)  \\  \label{HK9}
R_{U_1}+R_{V_1}  & ~~{\le}~~  I(U_1,V_1;Y_1|U_2)    \\ \label{HK10}
 R_{U_2}+R_{V_2}   & ~~{\le}~~   I(U_2,V_2;Y_2|U_1)   \\ \label{HK11}
R_{U_2}+R_{V_1}   & ~~{\le}~~  I(U_2,V_1;Y_1|U_1)   \\ \label{HK12}  
R_{U_1}+R_{V_2}    & ~~{\le}~~   I(U_1,V_2;Y_2|U_2)   \\  \label{HK13}
 R_{U_1}+R_{U_2}+ R_{V_1}  & ~~{\le}~~   I(U_1,U_2,V_1;Y_1)    \\ \label{HK14}
 R_{U_1}+R_{U_2} + R_{V_2}    & ~~{\le}~~   I(U_1,U_2,V_2;Y_2)   \\ \label{HK15}
 E(X_1^2)& ~~=~~   P_1  \\  \label{HK16}
 E(X_2^2)& ~~=~~   P_2.
\end{align}
 The above expressions specify the intersection of two multiple access channels, denoted as $\overline{M\!AC_1}$ with rate-tuple 
$(R^{(1)}_{U_1},R^{(1)}_{U_2},R^{(1)}_{V_1},R^{(1)}_{V_2})$ and 
$\overline{M\!AC_2}$ with rate-tuple 
$(R^{(2)}_{U_1},R^{(2)}_{U_2},R^{(2)}_{V_1},R^{(2)}_{V_2})$, when the intersection is projected on $(R^{(1)}_{U_1},R^{(1)}_{U_2},R^{(1)}_{V_1})$ to form MAC1 and on $(R^{(1)}_{U_1},R^{(1)}_{U_2},R^{(1)}_{V_2})$ to form MAC2. Hereafter, in dealing with rates in the projected regions, the superscripts will be dropped since rate values in the intersection of two regions will be the same in both regions, and  the use of distinguishing superscript is not needed.  In expressions \ref{HK1} to  \ref{HK14}, the superscripts are dropped since  $(R_{U_1},R_{U_2},R_{V_1},R_{V_2})$ correspond to rates in the intersection of projections.  

In expressions \ref{HK1} to \ref{HK14}, $\overline{M\!AC_1}$ is projected on sub-space $(R_{U_1},R_{U_2},R_{V_1})$, resulting in a region denoted as MAC1, and $\overline{M\!AC_2}$ is projected on sub-space $(R_{U_1},R_{U_2},R_{V_2})$, resulting in a region denoted as MAC2. Due to the nested structure of multiple-access channel, projections MAC1 and MAC2 are multiple-access channels in their respective sub-spaces.  It is clear that in the intersection of these two projections, the only sub-region with potentially non-zero volume (area) correspond to that of $(R_{U_1},R_{U_2})$. 

Constraint \ref{HK13} shows that the sum-rate in MAC1 is obtained by treating $V_2$ as noise.  Constraints \ref{HK14} shows that the sum-rate in MAC2 is obtained by treating $V_1$ as noise. These constraints impose a limit on respective sum-rates which play a key role in forming the shape of the intersection. Due to treating $V_1$ as noise in $Y_2$, and $V_2$ as noise in $Y_1$, some of the constraints such as \ref{HK11} and \ref{HK12} will be redundant. The same fact is expressed differently in Section 4 of \cite{HK1} (see Section 4 of \cite{HK1} for removal of redundant constraints). A different method for removing redundant constraint is based on noting that by treating $V_1$ as noise in MAC1, and $V_2$ as noise in MAC2, with Gaussian $V_1$ and $V_2$, the partial sum-rates, i.e., $R_{U_1}+R_{U_2}+R_{V_1}$ in MAC1 and $R_{U_1}+R_{U_2}+R_{V_2}$ in MAC2 are minimized. This is due to the fact the sum-rates in $\overline{M\!AC_1}$ and $\overline{M\!AC_2}$ do not depend on respective layering structures (depend only on power and channel gains~\cite{HK1}), and by having $V_2$ as the last layer in $\overline{M\!AC_1}$, the rate of $V_2$ in $\overline{M\!AC_1}$ is maximum (capacity of an AWGN channel~\cite{HK1}) and consequently, the partial sum-rate $R_{U_1}+R_{U_2}+R_{V_1}$ in MAC1 is minimized. A similar argument applies to $\overline{M\!AC_2}$, concluding  the partial sum-rate $R_{U_1}+R_{U_2}+R_{V_2}$ in MAC2 is minimized. 

Now let us form MAC1 and MAC2 with the minimum number of constraints in \ref{HK1} to \ref{HK14} being satisfied with equality (see Section 4 of \cite{HK1} for the corresponding equations). Let us try to force some of the remaining constraints to be satisfied with equality, and project the resulting rate-tuple on the sub-spaces of MAC1 or MAC2. As the (partial) sum-rates of MAC1/MAC2 are minimum possible, such projections either fall outside MAC1/MAC2, or on their respective boundaries. Let us consider the problem from the perspective of rate values which form the left hand sides of constraints in \ref{HK1} to \ref{HK14}. Falling outside MAC1/MAC2 means if the rate values on the left hand side of such constraints are computed from the equation set forming MAC1/MAC2, the corresponding left hand sides will becomes less than their respective mutual information terms on the right hand sides. This means such constraints are not violated and can be removed. 
See Remark 3 of \cite{HK1} explaining the reason that optimum power allocation may shift some of such points to the boundary of MAC1/MAC2, meaning that the corresponding constraints are satisfied with equality. 

having This means if the redundant constraints are solved with equality, the resulting rate-tuple when projected on the sub-space of MAC1 falls outside the MAC1 region, and likewise for MAC2. 

Now let us project $\overline{M\!AC_1}$ and $\overline{M\!AC_2}$ on the sup-space spanned by $(R_{V_1},R_{V_2})$, and also project  MAC1 and MAC2 on the same sup-space.  
Due to the nested structure of multiple-access channel, projections of $\overline{M\!AC_1}$ and $\overline{M\!AC_2}$ on sup-space  $(R_{V_1},R_{V_2})$ are two-dimensional multiple access channels, denoted as mac1 and mac2 hereafter. 
In forming these projections, constraint \ref{HK13} entails in $\overline{M\!AC_1}$, layer $V_2$ is the last layer to be successively decoded, since it acts as noise in formation of sum-rate in MAC1.  Likewise, constraint \ref{HK14} entails in $\overline{M\!AC_2}$, layer $V_1$ is the last layer to be successively decoded, since it acts as noise in formation of sum-rate in MAC2. 

Note that in forming mac1, projection of $\overline{M\!AC_1}$ on the sub-space $(R_{V_1},R_{V_2})$ is realized when treating $V_2$ as noise (same as in MAC1), and likewise, in forming mac2, projection of $\overline{M\!AC_2}$ on the sub-space $(R_{V_1},R_{V_2})$ is realized when treating $V_1$ as noise (same as in MAC2). As a result, in the intersection of mac1 and mac2, the upper corner point of mac1 overlaps with the lower corner point of mac2. This means intersection of mac1 and mac2 is a rectangular region. The corner point of this rectangular region, i.e., 
the upper corner point of mac1 overlapping with the lower corner point of mac2,  is the only point in the intersection of the projections of MAC1 and MAC2 on the sup-space $(R_{V_1},R_{V_2})$. 

The above arguments describe HK rate region in terms of the intersection of 
$\overline{M\!AC_1}$ and $\overline{M\!AC_2}$ in their corner regions where $V_2$ is the last Gaussian layer in the code-book for $\overline{M\!AC_1}$ and $V_1$ is the last Gaussian layer in the code-book for $\overline{M\!AC_2}$. Projection of this intersection forming MAC1 (and MAC2) have a sum-rate limited by \ref{HK13}, where $V_2$ acts as noise (and by \ref{HK14}, where $V_1$ acts as noise), respectively.

Since $\overline{M\!AC_1}$ and $\overline{M\!AC_2}$ are both optimized using Gaussian code-books, various projections formed by the HK constraints in \ref{HK1} to \ref{HK14} are also optimized using Gaussian code-books. On the other hand: 

\begin{enumerate}
\item It is known that HK constraints specify an achievable rate region for the interference channel. 

\item \cite{HK1} presents an achievable region for GIC based on moving step-by-step along the boundary, and for each point finding the optimum code-structure and its associated power allocation. This achievable region using Gaussian code-books coincides with the intersection of MAC1 and MAC2 as explained in the context of HK rate region.

\item \cite{HK1} Appendix  1 shows that using Gaussian code-books in MAC1/MAC2 region wherein Gaussian $V_2$/$V_1$ is, respectively, the last layer acting as noise, the capacity region of 2-users GIC cannot extend beyond the intersection of MAC1 and MAC2. 
\end{enumerate}
Using the above points, it follows that the optimally enlarged HK region using Gaussian code-books achieves the capacity region of the  2-users GIC.

\end{document}